\documentclass[aps,prd,twocolumn,showpacs,eqsecnum,amsmath,nofootinbib]{revtex4}

\usepackage{graphicx}

\begin{document}
\def \d {{\rm d}}
\def \A {{\rm A}}
\def \B {{\rm B}}
\def \CU {{\cal U}}
\def \CV {{\cal V}}

\title{Geodesics in spacetimes with expanding impulsive gravitational waves}

\author{Ji\v{r}\'{\i} Podolsk\'y}
\email{jiri.podolsky@mff.cuni.cz}
\affiliation{
  Institute of Theoretical Physics,  Charles University in Prague,\\
  V Hole\v{s}ovi\v{c}k\'{a}ch 2, 180 00 Prague 8, Czech Republic
  }
\author{Roland Steinbauer}
\email{roland.steinbauer@univie.ac.at}
\affiliation{
  Department of Mathematics, University of Vienna, Strudlhofg. 4,
  A-1090 Wien, Austria
  }

\date{\today}

\begin{abstract}
We study geodesic motion in expanding spherical impulsive gravitational
waves propagating in a Minkowski background.
Employing the continuous form of the metric we find and examine a
large family of geometrically preferred geodesics.
For the special class of axially symmetric spacetimes with the spherical
impulse generated by a snapping cosmic string we give a detailed
physical interpretation of the motion of test particles.

\end{abstract}

\pacs{04.20.Jb, 04.30.Nk}


\maketitle

\section{Introduction}\label{sec:intro}

In his classical work \cite{Pen72} Roger Penrose constructed impulsive
spherical gravitational waves in a Minkowski background using his
vivid ``cut and paste'' method. It is based on
cutting the spacetime along a null cone and then re-attaching the two
pieces with a suitable warp. An explicit solution using coordinates in
which the metric is continuous was later on given by Nutku and Penrose
\cite{NutPen92} and Hogan \cite{Hogan93,Hogan94}, but was only recently
related explicitly to the impulsive limit of
Robinson--Trautman type {\it N\,} solutions \cite{PodGri99,GriPodDoc02}.
However, the latter has to be considered
as only formal since the metric tensor contains terms proportional
to the square of the Dirac-$\delta$, and the transformation relating this
coordinate system to the continuous one mentioned above is necessarily discontinuous.
Nevertheless, this transformation is analogous to the one relating the distributional
and the continuous form of the metric tensor for impulsive {\em pp\,}-waves
(plane-fronted gravitational waves with parallel rays \cite{kramerbook}) which was
also introduced in \cite{Pen72} and has recently been analyzed {\it rigorously\,}
\cite{Stein98,KunSt99} using nonlinear theories of generalized functions
(Colombeau algebras) \cite{Col84,Col92}.
It is thus a natural open question whether a similar mathematically sound
treatment can also be found for expanding spherical impulses. This indeed
is one main motivation for the present work in which we study
the motion of test particles in spacetimes with spherical impulsive waves.

On the other hand this work is motivated by the quest of a physical
interpretation of radiative Robinson--Trautman spacetimes, one of the
most interesting non-static exact solutions of Einstein's
equations which admit a geodesic, shearfree and twistfree null
congruence of diverging rays \cite{kramerbook}. This large family
involves not only  spacetimes of Petrov type {\it N\,} (investigated
in the impulsive limit in the present paper) but also type {\it
II\,} solutions  describing bodies which radiate away their
asymmetries and approach a Schwarzschild black hole, or the
$C$-metric of type {\it D\,} which represents gravitational
radiation generated by uniformly accelerated black holes. By
studying these explicit exact solutions one may acquire an
intuition necessary for investigation of more general and
realistic situations.

This work is organized as follows. In Sec.\ \ref{sec:eiw} we
review the class of spacetimes under consideration and describe
the geometry of the expanding impulses.
By employing the continuous form of the metric in Sec.\ \ref{sec:geos} we find
a large class of privileged and simple geodesics which can be
related to explicit geodesics in the distributional form of
the metric ``in front'' and ``behind'' the spherical impulse.
This may allow one to lay the foundations for a
rigorous (distributional) treatment of impulsive Robinson--Trautman
solutions of type {\it N\,} as well as the transformation relating the
latter to the continuous form of the metric.
Moreover, assuming the geodesics to be $C^1$ across the impulse
(in the continuous system) we completely solve the problem of geodesic
motion in spacetimes with expanding impulsive gravitational waves.
In Sec.\ \ref{sec:string} we focus on impulsive waves generated by a snapping
cosmic string.  This interesting solution of Einstein's equations was
previously constructed by Gleiser and Pullin \cite{GlePul89} and Nutku and Penrose
\cite{NutPen92} using the ``cut and paste'' method. An independent
approach was used by Bi\v{c}\'ak \cite{Bicak90,BicSch89} (with recent
generalizations in \cite{PodGri01a}) who obtained the same spacetime
by considering a null limit of particular solutions with boost-rotational symmetry
representing a pair of particles uniformly accelerating due to semi-infinite strings
attached to them.
We discuss in detail the physical interpretation of the motion of test
particles influenced by such an impulse.

\section{Expanding impulsive waves in a Minkowski background}
\label{sec:eiw}

As mentioned above, Penrose \cite{Pen72} has described a ``cut and paste''
method for constructing expanding spherical gravitational waves in a
Minkowski background. The procedure can be performed explicitly
as follows. One starts with the Minkowski line element
 \begin{equation}
 \d s_0^2 = 2\,\d\eta\,\d\bar\eta  - 2\,\d \CU\,\d \CV
  = -\d t^2+\d x^2+\d y^2+\d z^2\,,
 \label{mink}
 \end{equation}
where the relation between the coordinates is given by
 \begin{equation}
  \CU=\textstyle{\frac{1}{\sqrt2}}(t+z)\ ,\
  \CV=\textstyle{\frac{1}{\sqrt2}}(t-z)\ ,\
  \eta=\textstyle{\frac{1}{\sqrt2}}(x+\,\hbox{i}\,y)\ .
 \label{trmink2}
 \end{equation}
We may now perform the transformation
 \begin{eqnarray}
 \CV&=& \frac{V}{p}-\epsilon U \ , \nonumber\\
 \CU&=& \frac{Z\bar Z}{p}\,V-U\ , \label{inv} \\
 \eta&=& \frac{Z}{p}\,V\ , \nonumber
 \end{eqnarray}
 where
 \begin{equation}
 p=1+\epsilon Z\bar Z\ , \qquad \epsilon=-1,0,+1\ .
 \label{pandep}
 \end{equation}
(The parameter $\epsilon$ is related to the Gaussian curvature of
the 2-surfaces given by $U=$~const.\, $V=$~const., cf.\ \cite{PodGri99}.)
Using (\ref{inv}), the metric (\ref{mink})
takes the form
 \begin{equation}
\d s_0^2 =
 2 \frac{V^2}{p^2}\,\d Z\,\d\bar Z +2\,\d U\,\d V -2\epsilon\,\d U^2\ .
 \label{U<0}
 \end{equation}
On the other hand, we consider the alternative, more involved transformation
given by
 \begin{eqnarray}
 \CV&=& AV-DU\ ,   \nonumber\\
 \CU&=& BV-EU\ ,   \label{transe3} \\
 \eta&=& CV-FU\ ,  \nonumber
 \end{eqnarray}
where
 \begin{eqnarray}
&&A= \frac{1}{p|h'|}\ ,\qquad
B= \frac{|h|^2}{p|h'|}\ ,\qquad
C= \frac{h}{ p|h'|}\ ,   \nonumber\\
&&D= \frac{1}{|h'|}\left\{
\frac{p}{4} \left|\frac{h''}{h'}\right|^2+\epsilon
\left[1+\frac{Z}{2}\frac{h''}{h'}+\frac{\bar Z}{2}\frac{\bar h''}{\bar h'}
\right]\right\}\ ,\nonumber\\
&&E= \frac{|h|^2}{|h'|}
\bigg\{ \frac{p}{4}\left|\frac{h''}{h'}-2\frac{h'}{h}\right|^2  \nonumber\\
&&\qquad+\epsilon\left[ 1+\frac{Z}{2}
\left(\frac{h''}{h'}-2\frac{h'}{h}\right)+\frac{\bar Z}{2}
\left(\frac{\bar h''}{\bar h'}-2\frac{\bar h'}{\bar h}\right)
\right]\bigg\}\ ,\nonumber\\
&&F= \frac{h}{|h'|}\bigg\{
\frac{p}{4}\left(\frac{h''}{h'}-2\frac{h'}{h}\right)
\frac{\bar h''}{\bar h'}  \nonumber\\
&&\qquad+\epsilon\left[1+
 \frac{Z}{2}\left(\frac{h''}{h'}-2\frac{h'}{h}\right)
+\frac{\bar Z}{2}\frac{\bar h''}{\bar h'}\right]\bigg\}\ .
 \label{transe4}
 \end{eqnarray}
Here $h\equiv h(Z)$ is an arbitrary function, and the derivative with respect to
its argument $Z$ is denoted by a prime. With this, the Minkowski metric
(\ref{mink}) becomes
 \begin{equation}
 \d s_0^2 = 2\left| \frac{V}{p}\,\d Z+U\,p\,\bar H\,\d\bar Z \right|^2
  +2\,\d U\,\d V -2\epsilon\,\d U^2\ ,
 \label{U>0}
 \end{equation}
 where $H$ is the Schwarzian derivative of $h$, i.e.,
 \begin{equation}
 H(Z)=\frac{1}{2} \left[\frac{h'''}{h'}-\frac{3}{2}\left(\frac{h''}{h'}\right)^2
 \right] .
 \label{Schwarz}
 \end{equation}
In the coordinates used in (\ref{U<0}) as well as in the ones used in (\ref{U>0}),
the null hypersurface \ $U=0$ \ represents a null cone
\hbox{$\eta\bar\eta-\CU\CV=0$}, i.e.,  an expanding sphere
$x^2+y^2+z^2=t^2$ in the Minkowski background. Moreover, the reduced
2-metrics on this cone are identical. Following the Penrose ``cut and paste''
method, we attach the line element (\ref{U<0}) for $U<0$ to (\ref{U>0}) for $U>0$.
The resulting metric takes the form
 \begin{equation}
\d s^2 =  2\left| \frac{V}{p}\,\d Z+U\Theta(U)\,p\,\bar H\,\d\bar Z
\right|^2\!\!+2\,\d U\,\d V -2\epsilon\,\d U^2\, ,
 \label{en0}
 \end{equation}
where $\Theta(U)$ is the Heaviside step function. This combined metric,
which was first presented in \cite{NutPen92,Hogan94},
is explicitly continuous everywhere, including the null
hypersurface $U=0$. However, the discontinuity in the derivatives of the
metric across $U=0$ yields an impulsive gravitational wave term proportional
to the Dirac $\delta$-function. More precisely, the only non-vanishing component of
the Weyl tensor in the Newman-Penrose formalism \cite{np, kramerbook} is
$\Psi_4=(p^2H/ V)\,\delta(U)$ which is the component
$\Psi_4\equiv C_{abcd}l^a\bar m^b l^c \bar m^d$ with respect to the null
tetrad ${\bf k}=\partial_V$, ${\bf l}=-\epsilon\,\partial_V-\partial_U$,
${\bf m}=p^2(V^2-U^2\Theta\, p^4 H\bar H)^{-1}[(V/p)\,\partial_{\bar Z}-U\Theta\, p\, \bar H\,\partial_Z]$.
The spacetime is thus flat everywhere except on the wave surface $U=0$.
Also, as shown in \cite{PodGri00}, the only non-vanishing tetrad
component of the Ricci tensor is $\Phi_{22}\equiv\frac12 R_{ab}l^a l^b=(p^4H\bar H/ V^2)\,U\delta(U)$.
This demonstrates that the spacetime is vacuum everywhere (except on the impulse at $V=0$
and at possible singularities of the function $p^2H$).

For later use we also remark that the inverse relation to (\ref{inv})
is given by
 \begin{eqnarray}
&&  U=\frac{\eta\bar\eta}{\CV}-\CU\ ,\nonumber\\
&&  V= \CV\ , \label{invepsis0}\\
&&  Z= \frac{\eta}{\CV}\ ,\nonumber
 \end{eqnarray}
when $\epsilon=0$, and by
 \begin{eqnarray}
&&  U= -\epsilon\CV-\frac{2\eta\bar\eta}
{(\CV-\epsilon\CU)\mp\sqrt{(\CV-\epsilon\CU)^2+4\epsilon\eta\bar\eta}}
  \ , \nonumber\\
&&  V= -(\CV-\epsilon\CU)-\frac{4\epsilon\eta\bar\eta}
{(\CV-\epsilon\CU)\mp\sqrt{(\CV-\epsilon\CU)^2+4\epsilon\eta\bar\eta}}\ ,
 \nonumber\\
&&  Z= -\frac{\epsilon}{2\bar\eta}\left[ (\CV-\epsilon\CU)
 \mp\sqrt{(\CV-\epsilon\CU)^2+4\epsilon\eta\bar\eta}\, \right]\ ,
 \label{invepsnot0}
 \end{eqnarray}
for $\epsilon\not=0$.

We may define the functions $U_{inv}$, $V_{inv}$ and  $Z_{inv}$ of $(U,V,Z,\bar Z)$
as the composition of (\ref{invepsis0}) (or (\ref{invepsnot0}) for
$\epsilon\not=0$) with (\ref{transe3}), (\ref{transe4}), which
transforms the metric (\ref{U>0}) to (\ref{U<0}).
Consequently, a discontinuous transformation
 \begin{eqnarray}
 u&=& U+\Theta(U)  \,[\, U_{inv}(U,V,Z,\bar Z)-U\,]\ , \nonumber\\
 w&=& V+\Theta(U)  \,[\, V_{inv}(U,V,Z,\bar Z)-V\,]\ , \label{transe}\\
 \xi&=& Z+\Theta(U)\,[\, Z_{inv}(U,V,Z,\bar Z)-Z \,]\ , \nonumber
 \end{eqnarray}
relates (\ref{en0}) for all $U\not=0$ to Minkowski spacetime in the
form
 \begin{equation}
\d s_0^2 =
 2 \frac{w^2}{\psi^2}\,\d\xi\,\d\bar\xi +2\,\d u\,\d w -2\epsilon\,\d u^2\ ,
 \label{backg}
 \end{equation}
where $\psi=1+\epsilon\xi\bar\xi$. Interestingly, by considering the
transformation (\ref{transe}) of the metric (\ref{en0}) for arbitrary $U$,
i.e., keeping also the distributional terms, one (formally) obtains the metric
  \begin{eqnarray}
\d s^2 &=& 2 \frac{w^2}{\psi^2}\left|\d\xi-f\,\delta(u)\d u\,\right|^2
+2\,\d u\,\d w -2\epsilon\,\d u^2 \nonumber\\
&&\!\!+w\Big[(f_\xi+\bar f_{\bar\xi}) -\frac{2\epsilon}{\psi}(f\bar\xi +\bar f\xi)
\Big] \delta(u) \d u^2\ , \label{RTN0e}
 \end{eqnarray}
in which $f(\xi)$ is related to $h(Z)$ through the identification
$f\equiv Z_{inv}-Z$ evaluated on $U=0$. Although this form of the metric
is only formal since it contains the square of the delta function, it
explicitly shows that expanding impulsive gravitational waves (\ref{en0})
arise as the impulsive limits of the
Robinson--Trautman type {\it N\,} spacetimes, expressed in the
coordinates introduced in \cite{GarPle81} (see \cite{BicPod99,PodGri99}
for the details).

Let us now conclude this review section by a brief description of
the geometry of the above expanding impulsive waves localized along the
wave-surfaces $U=0$, i.e., $u=0$. This will also elucidate the meaning
of the parameter $\epsilon$.

 From the Robinson--Trautman form of the metric
(\ref{RTN0e}) it is obvious that the
impulse  splits the spacetime into two flat regions, $u>0$ and $u<0$.
In the following we shall call the Minkowski half-space $u>0$ as being
``in front of the wave'', and the other Minkowski half-space $u<0$
(note that $u\equiv U$ for $U<0$) as being ``behind the wave''.
The ``background'' metric on both sides of the impulse is given by
(\ref{backg}). For {\it arbitrary\,} $u\not=0$, this metric can be
put into explicit Minkowski form (\ref{mink}) by the transformations (\ref{trmink2})
and (\ref{invepsis0}) (or  (\ref{invepsnot0})) and the
(trivial) identification $u=U$, $w=V$, $\xi=Z$. Using these relations,
we can easily analyze the geometry of the null hypersurfaces $u=u_0=$const.\
in Minkowski coordinates which are geometrically privileged and
thus allow for a clear physical interpretation.

We start with the subclass of solutions for which $\epsilon=0$.
Substituting (\ref{trmink2}) into (\ref{invepsis0}) and
setting $U=u_0$, we get the relation
\begin{equation}
x^2+y^2+(z+{\textstyle\frac{1}{\sqrt2}}\,u_0\,)^2=(t+{\textstyle\frac{1}{\sqrt2}}\,u_0\,)^2
\label{cone0}
\end{equation}
(if $t\not=z$, i.e., for $x\not=0$, $y\not=0$).
For various values of $u_0$ this represents a family of {\em null
cones} with vertices at $(-{\textstyle\frac{1}{\sqrt2}}\,u_0, 0, 0,
-{\textstyle\frac{1}{\sqrt2}}\,u_0\,)$
localized along a singular null line $t=z, x=0, y=0$.
Also, $V=w_0=\,$const.\ is a set of parallel hyperplanes
$t=z+\sqrt2\, w_0$ in the Minkowski
background. This reveals the geometrical meaning of the coordinates
$u, w$ used in the metric (\ref{backg}) with $\epsilon=0$.
Note that $w=0$ represents a physical singularity in the Robinson--Trautman
spacetimes \cite{BicPra98} which can be interpreted as the source of
the wave surfaces $u=u_0$. At any time $t$, these surfaces are
spheres of the radius $R=|t+{\textstyle\frac{1}{\sqrt2}}\,u_0|$. In particular, the
impulse localized on $u=0$ is a null cone with the vertex in
the origin which, at any time, is a sphere of radius
$R=\sqrt{x^2+y^2+z^2}=|t|$.

Analogous results can similarly be obtained for the remaining two
subclasses $\epsilon=\pm1$. In this case, using
(\ref{trmink2}) and (\ref{invepsnot0}), the hypersurfaces
$u_0$=const.\ are given by
\begin{equation}
x^2+y^2+z^2=(t+\sqrt2\,u_0)^2\ ,
\label{cone1}
\end{equation}
for $\epsilon=+1$ and
\begin{equation}
x^2+y^2+(z+\sqrt2\,u_0)^2=t^2\ ,
\label{cone-1}
\end{equation}
for $\epsilon=-1$, respectively. Again, these are families of null cones in
Minkowski space with vertices shifted in the $t$-direction for
$\epsilon=+1$, and in the $z$-direction for $\epsilon=-1$.
Note that these vertices form a singular timelike line $x=0=y$, $z=0$ if
$\epsilon=+1$, and a spacelike line $x=0=y$, $t=0$ if
$\epsilon=-1$. These lines are given by $\eta=0$,
$\CV-\epsilon\CU=0$, and correspond to the physical
singularity of the Robinson--Trautman spacetime at $w=0$.

It is obvious that for the above spacetimes with a {\it single} expanding impulsive wave
localized at $u=u_0=0$, i.e., at $U=0$, the null cones of the three classes of
wave-surfaces given by (\ref{cone0}), (\ref{cone1}), and (\ref{cone-1}) {\it coincide}.
In fact, in the impulsive limit the three (generically different)
subclasses $\epsilon=0,+1,-1$ of the Robinson--Trautman class of solutions
are {\it locally} equivalent \cite{PodGri99}.

It can also be observed, that for $\epsilon=0$, the physical singularity
at $V=0$ is a {\it singular null  line} on the wavefront surface $U=0$.
For a physical interpretation, it would be better to
remove this singularity from the spacetime. This can be achieved by
considering solutions with $\epsilon\ne0$: In these cases, the singularity at
$V=0$ appears {\it only at the vertex of} the null cone, $x=y=z=t=0$, which may be
considered as the ``origin'' of the spherical wave.

\section{Geodesic motion in spacetimes with expanding impulsive waves}
\label{sec:geos}

The purpose of this paper is to investigate the effect of
expanding impulsive waves of Robinson--Trautman type on the
motion of freely moving test particles. It is natural to start
with geodesics in (local, see below)
Minkowski space $U>0$ in front of the
wave, i.e., ``outside'' the null cone $U=0$ corresponding to
$x^2+y^2+z^2=t^2$. (Note that at $t=0$ the spherical impulse is
just ``created'' at the origin.) Obviously, general geodesics are
given by
\begin{eqnarray}
&&t^+=\gamma\,\tau\ ,  \nonumber\\
&&x^+=\dot x_0\,(\tau-\tau_i)+x_0\ , \label{geod+}\\
&&y^+=\dot y_0\,(\tau-\tau_i)+y_0\ , \nonumber\\
&&z^+=\dot z_0\,(\tau-\tau_i)+z_0\ , \nonumber
\end{eqnarray}
with $\gamma=\sqrt{\dot x_0^2+\dot y_0^2+\dot z_0^2-e}$, i.e.,
$\tau$ is a normalized affine parameter of timelike ($e=-1$) or
spacelike ($e=+1$) geodesics. For null geodesics ($e=0$) it is
always possible to scale the factor $\gamma$ to unity.  The
constants $x_0,y_0,z_0$ and  $\dot x_0,\dot y_0, \dot z_0$
characterize position resp.\ velocity of each test particle at the instant
\begin{equation}
\tau_i=\sqrt{x_0^2+y_0^2+z_0^2}\,/\,\gamma\ , \label{taui}
\end{equation}
when the geodesic intersects the null cone. At $\tau_i$ each
particle is hit by the impulse and its trajectory is refracted and
(possibly) shifted. The geodesics (\ref{geod+}) in front of the impulse
can also be written as
\begin{eqnarray}
\CV^+ &=&\dot\CV_0^+(\tau-\tau_i)+\CV_0^+\ ,\nonumber\\
\CU^+ &=&\dot\CU_0^+(\tau-\tau_i)+\CU_0^+\ ,\label{geod++}\\
\eta^+&=&\dot\eta_0^+(\tau-\tau_i)+\eta_0^+\ ,\nonumber
\end{eqnarray}
where
\begin{eqnarray}
&&\dot\CV_0^+=\textstyle{\frac{1}{\sqrt2}}(\gamma-\dot z_0)\ ,
\quad\
   \CV_0^+=\textstyle{\frac{1}{\sqrt2}}(\gamma\tau_i-z_0)\ ,\nonumber\\
&&\dot\CU_0^+=\textstyle{\frac{1}{\sqrt2}}(\gamma+\dot z_0) \ ,
\quad\
   \CU_0^+=\textstyle{\frac{1}{\sqrt2}}(\gamma\tau_i+z_0)\ ,\label{coef+}\\
   &&\dot\eta_0^+=\textstyle{\frac{1}{\sqrt2}}(\dot x_0+\hbox{i}\, \dot y_0)\ , \quad
   \eta_0^+=\textstyle{\frac{1}{\sqrt2}}(x_0+\hbox{i}\,y_0)\ .\nonumber
\end{eqnarray}

Now, we wish to investigate the influence of the impulse on the geodesics, and
to determine explicitly formulae for corresponding ``refraction'' and
``shift''. In the region $U<0$ behind the impulse the particles again
move in Minkowski space (\ref{mink}) so that the trajectories
also have to be straight lines of the form
\begin{eqnarray}
\CV^- &=&\dot\CV_0^-(\tau-\tau_i)+\CV_0^-\ , \nonumber\\
\CU^- &=&\dot\CU_0^-(\tau-\tau_i)+\CU_0^-\ ,\label{geod-}\\
\eta^-&=&\dot\eta_0^-(\tau-\tau_i)+\eta_0^-\ .\nonumber
\end{eqnarray}
It remains to express the constants appearing in (\ref{geod-})
in terms of the initial data introduced in (\ref{geod+}), and the
structural function $h(Z)$, which characterizes specific expanding
impulsive waves.

The key idea is to {\em employ the continuous form} of the
solution (\ref{en0}). It can easily be observed that in this
coordinate system the Christoffel symbols $\,\Gamma^\mu_{UU}$,
$\Gamma^\mu_{UV}$, and $\Gamma^\mu_{VV}$ vanish identically.
Therefore, the geodesic equations {\it always admit privileged
global solutions} of the form
\begin{eqnarray}
Z&=&Z_0=\hbox{const.}\ ,\nonumber\\
U&=&\dot U_0(\tau-\tau_i)\ ,\label{privgeod}\\
V&=&\dot V_0(\tau-\tau_i)+V_0\ .\nonumber
\end{eqnarray}
Here we have set $U_0=0$, so that each geodesic
reaches the impulse localized at $U=0$ at parameter-time
$\tau=\tau_i$.

Using (\ref{transe3}), it is  possible to express the 
geodesics (\ref{privgeod}) {\it in front} of
the impulse in the Minkowski form (\ref{geod++}) where the coefficients
(\ref{coef+}) are
\begin{eqnarray}
&&\dot\CV_0^+=A\dot V_0-D \dot U_0\ , \quad
   \CV_0^+=AV_0\ ,\nonumber\\
&&\dot\CU_0^+=B\dot V_0-E \dot U_0\ , \quad
   \CU_0^+=BV_0\ ,\label{coef++}\\
&&\dot\eta_0^+\,=C\dot V_0-F \dot U_0 \ , \quad\>
   \eta_0^+=\,CV_0\ .\nonumber
\end{eqnarray}
The constants $A,B,C,D,E$ and $F$ are given by the values of the functions
(\ref{transe4}) at $Z=Z_0$. The relations (\ref{coef+})
and (\ref{coef++}) enable us to relate the parameters $Z_0$, $\dot U_0$,
$\dot V_0$, and $V_0$ to the natural initial data introduced in
(\ref{geod+}) by
\begin{eqnarray}
&&x_0=\sqrt2 \,\, V_0\,{\cal R}e\, C\ ,\nonumber\\
&&y_0=\sqrt2 \,\, V_0\,{\cal I}m\, C\ ,\nonumber\\
&&z_0=\textstyle\frac{1}{\sqrt2}\,V_0\,(B-A)\ ,\label{relation}\\
&&\dot x_0=\sqrt2 \,\,[\, \dot V_0\,{\cal R}e\, C-\dot U_0\,{\cal R}e\, F\,]\ ,\nonumber\\
&&\dot y_0=\sqrt2 \,\,[\, \dot V_0\,{\cal I}m\, C-\dot U_0\,{\cal I}m\, F\,]\ ,\nonumber\\
&&\dot z_0=\textstyle\frac{1}{\sqrt2}\,[\, \dot V_0\,(B-A)-\dot U_0\,(E-D)\,]\ .\nonumber
\end{eqnarray}
The three position parameters $x_0$, $y_0$, $z_0$ are thus related
to the three independent constants $Z_0$, $\bar Z_0$, $V_0$.
The normalization condition 
implies the constraint
$(\dot V_0-\epsilon\dot U_0)\dot U_0={\textstyle \frac{1}{2}}e$
(which can be obtained using the identities $AB-C\bar C=0$,
$DE-F\bar F=\epsilon$, and $AE+BD-C\bar F-\bar C F=1$)
on the parameters $\dot V_0$ and $\dot U_0$. However,
in general it is possible to set at least one of the velocities
$\dot x_0$, $\dot y_0$, or $\dot z_0$ to zero by a suitable choice
of $\dot U_0$.

Finally, we transform the geodesics (\ref{privgeod}) {\it
behind} the impulse using (\ref{inv}), which gives the uniform Minkowskian
motion (\ref{geod-}) with the coefficients
\begin{eqnarray}
&&\dot\CV_0^-=\frac{\dot V_0}{p}-\epsilon\dot U_0 \ , \quad\qquad
   \CV_0^-=\frac{V_0}{p}\ ,\nonumber\\
&&\dot\CU_0^-=\frac{Z_0\bar Z_0}{ p}\dot V_0-\dot U_0\ , \quad
   \CU_0^-=\frac{Z_0\bar Z_0}{p}V_0\ ,\label{coef-}\\
&&\dot\eta_0^-=\frac{Z_0}{ p}\dot V_0 \ , \qquad\qquad\quad
   \eta_0^-=\frac{Z_0}{ p } V_0\ .\nonumber
\end{eqnarray}
Substituting for $Z_0$, $V_0$, $\dot V_0$ and $\dot U_0$ from
the inverse of the relations (\ref{relation}), we  obtain an explicit
result which can be used for discussion of the effect of expanding
impulsive waves on the privileged family of geodesics
(\ref{privgeod}).

However, it should be emphasized that in the above construction we
started with the initial data (\ref{geod+}) {\it outside} the impulse
in the region $U>0$, which is a Minkowski space {\it only
locally}. Due to the complicated form of the generating complex
function $h$, there are topological defects such as cosmic strings
(corresponding to the presence of deficit angles) outside the null
cone, see e.g., \cite{PodGri00} for more details. Therefore, for better
physical interpretation it may be useful to set the initial data for
the geodesics in the region $U<0$ {\it inside} the cone, which is
considered to be a ``complete'' Minkowski space without topological
defects. Evolving these data,
\begin{eqnarray}
&&t^-=\gamma\,\tau\ ,  \nonumber\\
&&x^-=\dot x_0\,(\tau-\tau_i)+x_0\ , \label{geod+b}\\
&&y^-=\dot y_0\,(\tau-\tau_i)+y_0\ , \nonumber\\
&&z^-=\dot z_0\,(\tau-\tau_i)+z_0\ , \nonumber
\end{eqnarray}
``backward'' in time (i.e., for decreasing affine parameter
$\tau$), it is possible to prolong the geodesics across the
spherical impulse to  the ``incomplete'' Minkowski region $U>0$
with cosmic strings outside the impulse.

In this case, the {\it explicit  geodesics outside}
the impulse have the form (\ref{geod++}), (\ref{coef++}), in which the constants
$Z_0$, $V_0$, $\dot V_0$ and $\dot U_0$ have to be expressed in terms
of the data (\ref{geod+b}). Using (\ref{invepsis0}) for
$\epsilon=0$, and (\ref{invepsnot0}) for $\epsilon\not=0$ we obtain
\begin{eqnarray}
&&Z_0=\frac{x_0 +\, \hbox{i}\, y_0}{\gamma\tau_i-z_0}\ ,\nonumber\\
&&V_0={\textstyle\frac{1}{\sqrt2}}\,[(1+\epsilon)\gamma\tau_i-(1-\epsilon)
z_0]\ ,\nonumber\\
&& \dot V_0={\textstyle\frac{1}{\sqrt2}}\,[(1-\epsilon)\gamma-
(1+\epsilon)\dot z_0]\, \frac{(1+\epsilon)\gamma\tau_i-(1-\epsilon)
z_0} {(1-\epsilon) \gamma\tau_i-(1+\epsilon)z_0}\
,\nonumber\\
&&\dot U_0=\frac{\sqrt2\,\gamma(z_0-\tau_i\dot z_0)}
{(1-\epsilon)\gamma\tau_i-(1+\epsilon) z_0}\ ,\label{cone}
\end{eqnarray}
with the constraint
\begin{eqnarray}
&& [(1-\epsilon)\gamma\tau_i-(1+\epsilon)z_0)]\,
  (\dot x_0 + \,\hbox{i}\, \dot y_0)   \label{constraint}\\
&&\hskip15mm =[(1-\epsilon)\gamma-(1+\epsilon)\dot z_0]\,
  (x_0 + \,\hbox{i}\, y_0)\ . \nonumber
\end{eqnarray}
In (\ref{cone}) above we assumed that $(1-\epsilon)\gamma\tau_i\not=(1+\epsilon)z_0$.
The case $(1-\epsilon)\gamma\tau_i=(1+\epsilon)z_0$ requires $x_0=y_0=z_0=0$.
Such geodesics reach the singular vertex $x=y=z=0$ of the impulsive null
cone at $t=\tau=0$, and it is thus unphysical to investigate their
continuation across $U=0$. On the other hand, if
$(1-\epsilon)\gamma=(1+\epsilon)\dot z_0$ then $\dot x_0=\dot y_0=0$, and
$\dot V_0=0$. These geodesics are {\it null} for $\epsilon=0$
(with $\dot U_0=-{\textstyle{\sqrt2}}\,\dot z_0$), {\it timelike} for
$\epsilon=1$ (with $\gamma=1$, $\dot z_0=0$, $\dot U_0=-1/\sqrt2$), and
{\it spacelike} for $\epsilon=-1$.

Note that the relations (\ref{cone}) simplify for each particular choice of
the parameter $\epsilon$. In particular, in the case $\epsilon=0$ we obtain
\begin{eqnarray}
&&Z_0=\frac{x_0 +\, \hbox{i}\, y_0}{\gamma\tau_i-z_0}\ ,\qquad
V_0={\textstyle\frac{1}{\sqrt2}}\,(\gamma\tau_i-z_0)\ ,\nonumber\\
&&\dot V_0={\textstyle\frac{1}{\sqrt2}}\,(\gamma-\dot z_0)\ ,\quad\>
\dot U_0={\textstyle\frac{1}{\sqrt2}}\,\frac{e}{\gamma-\dot z_0}
\ .\label{con}
\end{eqnarray}

Let us finally recall that the above class of geodesics is
privileged and very special since $Z=Z_0=\hbox{const.}$ (cf.\ (\ref{privgeod})).
However, it is possible to find {\it general\,} geodesics
{\it assuming them to be $C^1$ across the impulse}
in the continuous coordinate system (\ref{en0}).
With this assumption, the constants
\begin{eqnarray}
&&Z_i\equiv Z(\tau_i)\ ,\quad V_i\equiv V(\tau_i)\ ,\quad
  U_i\equiv U(\tau_i)=0\ ,\nonumber\\
&&\dot Z_i\equiv \dot Z(\tau_i)\ ,\quad\dot V_i\equiv \dot V(\tau_i)\ ,\quad
  \dot U_i\equiv \dot U(\tau_i)\ ,
\label{const}
\end{eqnarray}
describing positions and velocities at $\tau_i$, the instant of
interaction with the impulsive wave, have the {\it same} values when evaluated
in the limits $U\to0$ both from the region in front $(U>0)$ and behind
the impulse $(U<0)$. Starting now with the general initial data (\ref{geod+b})
in the region $U<0$, in which (\ref{invepsis0}), (\ref{invepsnot0})
apply, it is straightforward to derive
\begin{eqnarray}
&&Z_i=\frac{x_0 +\, \hbox{i}\, y_0}{\gamma\tau_i-z_0}\ ,\nonumber\\
&&V_i={\textstyle\frac{1}{\sqrt2}}\,[(1+\epsilon)\gamma\tau_i-(1-\epsilon)
z_0]\ ,\nonumber\\
&&\dot Z_i=\frac{\dot x_0 +\, \hbox{i}\, \dot y_0}{\gamma\tau_i-z_0}
-\frac{x_0 +\, \hbox{i}\, y_0}{(\gamma\tau_i-z_0)^2}\nonumber\\
&& \,\times\frac{[(1-\epsilon)\gamma-(1+\epsilon)\dot z_0](\gamma\tau_i-z_0)
  +2\epsilon(x_0\dot x_0+y_0\dot y_0)}
  {(1+\epsilon) \gamma\tau_i-(1-\epsilon)z_0} \ ,\nonumber\\
&& \dot V_i={\textstyle\frac{1}{\sqrt2}}\Big\{
[(1-\epsilon)\gamma- (1+\epsilon)\dot z_0]\,
[(1-\epsilon)\gamma\tau_i-(1+\epsilon)z_0]  \nonumber\\
&&\hskip8mm +4\epsilon(x_0\dot x_0+y_0\dot y_0)\Big\}\,
  [(1+\epsilon) \gamma\tau_i-(1-\epsilon)z_0]^{-1}\ ,\nonumber\\
&&\dot U_i={\textstyle{\sqrt2}}\,\,
\frac{x_0\dot x_0+y_0\dot y_0+z_0\dot z_0-\gamma^2\tau_i}
{(1+\epsilon)\gamma\tau_i-(1-\epsilon) z_0}\ .\label{genercont}
\end{eqnarray}
Then, using (\ref{transe3}) the  geodesics outside the impulse
in the (local) Minkowski space $U>0$ are given by the
expressions (\ref{geod++}) with
\begin{eqnarray}
&& \CV_0^+=AV_i\ ,\nonumber\\
&& \CU_0^+=BV_i\ ,\nonumber\\
&& \eta_0^+\,=\,CV_i\ ,\nonumber\\
&&\dot\CV_0^+=A\dot V_i-D \dot U_i
   +(A_{,Z}\dot Z_i+A_{,\bar Z}\dot{\bar Z}_i) V_i\ ,\label{complcoef++}\\
&&\dot\CU_0^+=B\dot V_i-E \dot U_i
   +(B_{,Z}\dot Z_i+B_{,\bar Z}\dot{\bar Z}_i) V_i\ , \nonumber\\
&&\dot\eta_0^+\,=C\dot V_i-F \dot U_i
   +(C_{,Z}\dot Z_i+C_{,\bar Z}\dot{\bar Z}_i) V_i \ ,\nonumber
\end{eqnarray}
in which the coefficients  and their derivatives are given by the
functions (\ref{transe4}), evaluated at $Z_i$.

Of course, with the constraint (\ref{constraint}) we obtain $\dot Z_i=0$, $Z_i=Z_0$,
$V_i=V_0$, $\dot V_i=\dot V_0$, $\dot U_i=\dot U_0$, and the above
geodesics reduce to the privileged family presented in (\ref{cone}).

\section{Geodesics in spacetimes with a snapping cosmic string}
\label{sec:string}

In this section we apply the above general results to an interesting particular
class of spacetimes in which the expanding spherical impulsive wave is generated
by a snapping cosmic string (identified outside the impulse by a deficit angle). This solution
was previously introduced and discussed in a number of  works
\cite{NutPen92,GlePul89,Bicak90,PodGri01a}.
It can be written as the metric (\ref{en0}) with
\begin{equation}
H(Z)=\frac{\frac{1}{2}\delta(1-\frac{1}{2}\delta)}{Z^2}\ , \label{string}
\end{equation}
which is generated by (\ref{Schwarz}) with $h(Z)=Z^{1-\delta}$.
Here $\delta$ is a real positive constant, $\delta <1$ which characterizes the
deficit angle $2\pi\delta$ of the snapped string localized outside the impulse
along the axis $\eta=0$ (see, e.g. \cite{NutPen92,PodGri00}).
It is straightforward  to calculate the coefficients
(\ref{transe4}) and their derivatives for such 
a function $h$, i.e.,
 \begin{eqnarray}
&&A= \frac{|Z|^\delta}{(1-\delta)p}\ ,\ \quad
B= \frac{|Z|^{2-\delta}}{(1-\delta)p}\ ,\ \quad
C= \frac{Z^{1-\delta} |Z|^\delta}{(1-\delta)p}\ ,   \nonumber\\
&&D= \frac{|Z|^{\delta-2}}{1-\delta}\left[ (\textstyle{\frac{1}{2}\delta})^2
    +(\textstyle{1-\frac{1}{2}\delta})^2\epsilon |Z|^2\right]\ ,\nonumber\\
&&E= \frac{|Z|^{-\delta}}{1-\delta}\left[ (\textstyle{1-\frac{1}{2}\delta})^2
    +(\textstyle {\frac{1}{2}\delta})^2\epsilon |Z|^2\right]\ ,\nonumber\\
&&F= \frac{Z^{1-\delta}|Z|^{\delta-2} \textstyle{ \frac{1}{2}\delta(1-\frac{1}{2}\delta)p}}
       {1-\delta}\ ,\label{ABCDEFgener}\\
&&A_{,Z}= \frac{|Z|^\delta[\textstyle{\frac{1}{2}\delta}
    -(\textstyle{1-\frac{1}{2}\delta})\epsilon|Z|^2]}{(1-\delta)p^2Z}\ ,\nonumber\\
&&B_{,Z}= \frac{|Z|^{2-\delta}[(\textstyle{1-\frac{1}{2}\delta})
    -\textstyle{\frac{1}{2}\delta}\epsilon|Z|^2]}{(1-\delta)p^2Z}\ ,\nonumber\\
&&C_{,Z}=  \left(\frac{\bar Z}{ Z}\right)^{\delta/2}\frac{(\textstyle{1-\frac{1}{2}\delta})
    -\textstyle{\frac{1}{2}\delta}\epsilon|Z|^2}{(1-\delta)p^2}\ ,\nonumber\\
&&C_{,\bar Z}=  \left(\frac{\bar Z}{ Z}\right)^{\delta/2-1}
  \frac{\textstyle{\frac{1}{2}\delta}-(\textstyle{1-\frac{1}{2}\delta})\epsilon|Z|^2}
    {(1-\delta)p^2}    \ .\nonumber
 \end{eqnarray}

\subsection{Privileged geodesics with $Z=Z_0$ }

We first consider the family of geometrically preffered geodesics
(\ref{privgeod}) for which $Z=Z_0=\,$const.\ It is convenient to use the global
axial symmetry of the solution (\ref{en0}), (\ref{string}) corresponding to a coordinate freedom
$Z\to Z\,\exp(\hbox{i}\phi)$, where $\phi$ is a constant. Therefore, without loss of generality we
can assume $Z_0$ to be a {\it real} positive constant, in which case the coefficients
(\ref{ABCDEFgener}) reduce to
 \begin{eqnarray}
&&A= \frac{Z_0^\delta}{(1-\delta)p}\ ,\ \quad
B= \frac{Z_0^{2-\delta}}{(1-\delta)p}\ ,\ \quad
C= \frac{Z_0}{(1-\delta)p}\ ,   \nonumber\\
&&D= \frac{Z_0^{\delta-2}}{1-\delta}\left[ (\textstyle{\frac{1}{2}\delta})^2
    +(\textstyle{1-\frac{1}{2}\delta})^2\epsilon Z_0^2\right]\ ,\nonumber\\
&&E= \frac{Z_0^{-\delta}}{1-\delta}\left[ (\textstyle{1-\frac{1}{2}\delta})^2
    +(\textstyle {\frac{1}{2}\delta})^2\epsilon Z_0^2\right]\ ,\label{ABCDEF}\\
&&F= \frac{ \textstyle{ \frac{1}{2}\delta(1-\frac{1}{2}\delta)p}} {(1-\delta)Z_0}\ ,\nonumber
 \end{eqnarray}
where $\,p=1+\epsilon Z_0^2$. Substituting 
(\ref{ABCDEF}) into (\ref{relation}), we obtain the relations
between the parameters $Z_0$, $\dot U_0$, $\dot V_0$  and $V_0$
characterizing the family of geodesics (\ref{privgeod}) and the
initial data {\it outside} the spherical impulse (cf. (\ref{geod+})).
Obviously, these geodesics all have $y^+\equiv 0$, since
$\dot y_0=0=y_0$ due to vanishing imaginary parts of $C$ and $F$.
The remaining relations (\ref{relation}) yield
 \begin{eqnarray}
 Z_0^{1-\delta} &=& \frac{z_0}{x_0}+\sqrt{1+\frac{z_0^2}{x_0^2}}\ ,   \label{inver}\\
 \dot U_0 &=& \sqrt2\frac{z_0\dot x_0-x_0\dot z_0}{(E-D)\,x_0-2F\,z_0}\ ,\nonumber\\
\frac{Z_0}{p} V_0&=& {\textstyle\frac{1}{\sqrt2}}(1-\delta)\,x_0\ ,\nonumber\\
\frac{Z_0}{p}\dot V_0&=& {\textstyle\frac{1}{\sqrt2}}(1-\delta)\,x_0
   \frac{(E-D)\,\dot x_0-2F\,\dot z_0}{(E-D)\,x_0-2F\,z_0}\ .\nonumber
 \end{eqnarray}
(Note that ``initially static'' observers $\dot x_0=0=\dot z_0$ are excluded from the
family (\ref{privgeod}) as this would give $\dot U_0=0=\dot V_0$, i.e., the  geodesic
is constant.) Finally, substituting (\ref{inver}) into (\ref{coef-}) we obtain
\begin{eqnarray}
x_0^-&=&(1-\delta)\,x_0\ ,\nonumber\\
y_0^-&=&0\ ,\nonumber\\
z_0^-&=&{\textstyle\frac{1}{2}}(1-\delta)\,x_0(Z_0-Z_0^{-1}) \ ,\nonumber\\
t_0^-&=&{\textstyle\frac{1}{2}}(1-\delta)\,x_0(Z_0+Z_0^{-1}) \ ,\nonumber\\
\dot x_0^-&=&(1-\delta)\,x_0\frac{(E-D)\,\dot x_0-2F\,\dot z_0}{(E-D)\,x_0-2F\,z_0}\ ,
\label{explicit}\\
\dot y_0^-&=&0\ ,\nonumber\\
\dot z_0^-&=&\frac{z_0^-[(E-D)\,\dot x_0-2F\,\dot z_0]
  -(1-\epsilon)(z_0\dot x_0-x_0\dot z_0)}{(E-D)\,x_0-2F\,z_0} \ ,\nonumber\\
\dot t_0^-&=&\frac{t_0^-[(E-D)\,\dot x_0-2F\,\dot z_0]
  -(1+\epsilon)(z_0\dot x_0-x_0\dot z_0)}{(E-D)\,x_0-2F\,z_0} \ ,\nonumber
\end{eqnarray}
where the coefficients $E-D$ and $F$ are
\begin{eqnarray}
(1-\delta)\,(E-D)&=& (\textstyle{1-\frac{1}{2}\delta})^2
  (Z_0^{-\delta}-\epsilon Z_0^\delta)\nonumber\\
  &&\quad-  (\textstyle {\frac{1}{2}\delta})^2(Z_0^{\delta-2}-\epsilon
  Z_0^{2-\delta})\ ,\nonumber\\
(1-\delta)\,2F&=& \delta(\textstyle{1-\frac{1}{2}\delta})(Z_0^{-1}+\epsilon Z_0)\ ,
\label{EDF}
\end{eqnarray}
and $Z_0$ is given by (\ref{inver}).

The above relations explicitly express the effect of an expanding impulsive wave generated
by a snapping cosmic string on  geodesics (\ref{geod+}) of the privileged family
(\ref{privgeod}). These start outside the impulse ($U>0$) with the initial
data entering the right hand side of (\ref{explicit}), and continue
inside the spherical impulse in the Minkowski space without the
string ($U<0$), where the positions and velocities at $\tau=\tau_i$ are now given by the
left hand side of (\ref{explicit}). For $\delta=0$ we
obtain $E-D=1-\epsilon$, $F=0$, $Z_0-Z_0^{-1}=2z_0/x_0$,
$Z_0+Z_0^{-1}=2\gamma\tau_i/x_0$, so that $x^-=x^+$, $y^-=0=y^+$,
$z^-=z^+$, $t_0^-=\gamma\tau_i$, and $\dot t_0^-=\gamma$ (to
derive the last relation  we have used  the constraint
(\ref{constraint})). Obviously, these are {\it global} geodesics
in {\it complete} Minkowski space without the strings and the impulse.

Let us now investigate in some more detail the effect of the spherical gravitational
impulse on free test particles.
To describe the ``refraction'' and the ``shift'' of geodesic trajectories it is convenient
to introduce angles $\alpha$ and $\beta$, whose geometrical meaning is indicated in Fig.~1.
Recall that for the special family of geodesics (\ref{privgeod}) we have $y^-=0=y^+$,
so that the motion is confinded to $x,z$-plane which contains the string (located along the
$z$-axis). Hence $\alpha$ and $\beta$
represent the {\it position} of the particle
at the instant $\tau_i$ of interaction with the impulse resp.\ the direction of its
{\it velocity} (inclination of the trajectory) in the $x,z$-plane. 

\begin{figure}
\centering
\includegraphics[width=.47\textwidth]{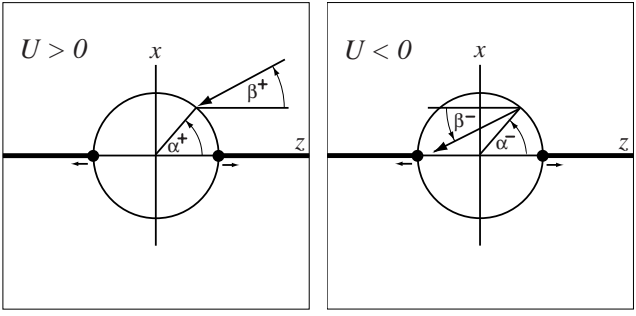}
\vspace*{20pt}
\caption{\label{fig1}
The angle $\alpha$ identifies the point where the particle
interacts with the impulse (indicated by a circle) generated by a
snapped string localized on the $z$-axis. The angle $\beta$
characterizes the inclination of its trajectory. The superscripts
``+'' and ``--'' correspond to the respective
values in the (local) Minkowskian coordinate system outside ($U>0$) and
the (different) Minkowskian system inside ($U<0$) the impulse.
}%
\end{figure}

\medskip
\noindent
In the region $U>0$ outside the impulsive wave these parameters
are defined as
\begin{equation}
\cot\alpha^+={z_0/x_0}\ ,\qquad \cot\beta^+={\dot z_0/\dot x_0}\ .\label{ab+}
\end{equation}
Similarly, behind the impulse in the region $U<0$ 
we have
\begin{equation}
\cot\alpha^-={z_0^-/x_0^-}\ ,\qquad \cot\beta^-={\dot z_0^-/\dot x_0^-}\ .\label{ab-}
\end{equation}
Straightforward calculations using (\ref{inver}) give
\begin{equation}
Z_0^{1-\delta}=\cot\,({\textstyle\frac{1}{2}}\alpha^+)\ ,\label{Z0}
\end{equation}
which implies the relation
$\frac{1}{2}(Z_0^{1-\delta}-Z_0^{\delta-1})=\cot\alpha^+$. From
(\ref{explicit}) and (\ref{Z0}) we immediately obtain
\begin{eqnarray}
\cot\alpha^-&=&{\textstyle\frac{1}{2}}(Z_0-Z_0^{-1})\nonumber \\
&=&{\textstyle\frac{1}{2}}[\,\cot^{\,q}\,({\textstyle\frac{1}{2}}\alpha^+)
-\cot^{-q}\,({\textstyle\frac{1}{2}}\alpha^+)\,]\,, \label{alpha}
\end{eqnarray}
where $q=1/(1-\delta)$. This expression gives the relation
$\alpha^-(\alpha^+)$ which {\it identifies the points} on both sides of
the impulse in the natural Minkowskian coordinate systems.

Analogously we derive the following relation for the velocities,
\begin{equation}
\cot\beta^- -\cot\alpha^-={\cal N}\,(\cot\beta^+ -\cot\alpha^+)\ ,\label{beta}
\end{equation}
where
\begin{equation}
{\cal N}={\cal N}(\alpha^+,\beta^+)
        =\frac{1-\epsilon}{(1-\delta)(E-D)-(1-\delta)\,2F\cot\beta^+}\ .\label{N}
\end{equation}
This is the {\it refraction formula} for trajectories of free test
particles which cross the spherical impulse.

Notice that the above considerations also apply to geodesics propagating in
the privileged directions $\beta^+=0$ and $\beta^+=\frac{\pi}{2}$.
For geodesics {\it parallel to the string}, i.e., in the case $\beta^+=0$
(implying $\dot x_0=0$) the right hand side of (\ref{beta}) has to be replaced by
the simple expression $(\epsilon-1)/[(1-\delta)2F]$.
For trajectories with $\beta^+=\frac{\pi}{2}$ which are {\it perpendicular}
to the strings ($\dot z_0=0$), the right hand side simplifies to
$[(\epsilon-1)\cot\alpha^+]/[(1-\delta)(E-D)]$.

Several interesting observations can immediately be done. For
$\delta=0$ representing a complete Minkowski space without the impulse and
topological defects, one obtains $\alpha^-=\alpha^+$,  ${\cal N}=1$, and
consequently $\beta^-=\beta^+$. There is thus no ``shift'' and
``refraction'', as expected.

For a general $\delta$, it follows from (\ref{beta}) that if
$\alpha^+=\beta^+$ then $\alpha^-=\beta^-$.  This  means physically
that the {\it radial geodesics} (``perpendicular'' to the spherical impulse)
{\it remain radial} also behind the impulse.

Moreover, it can be observed from ({\ref{N}) that the coefficient ${\cal N}$
identically vanishes for spacetimes {\it with the parameter} $\epsilon=+1$.
Consequently, $\alpha^-=\beta^-$, which means that the  geodesics
(\ref{privgeod})  are refracted by the impulse in such a way that their
{\it trajectories become radial}. These
are thus either exactly focused towards the origin $x=0=z$, or defocused
directly  from it.

\begin{figure}
\includegraphics[width=.47\textwidth]{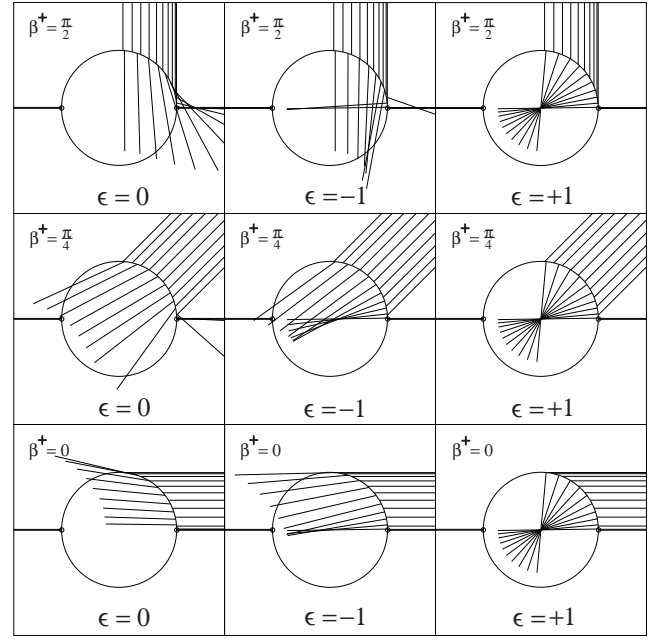}
\caption{\label{fig2}
Typical behavior of geodesic trajectories, which are
refracted and shifted by the expanding spherical impulse, for
various values of $\alpha^+$, $\beta^+$ and the parameter
$\epsilon$. Here $\delta=0.3$.
}%
\end{figure}

The typical behavior of geodesics affected by the impulsive gravitational wave
(\ref{en0}), (\ref{string}), as described by the refraction formula
(\ref{beta}), is shown in Fig.~2 for various choices of
$\alpha^+, \beta^+$, and $\epsilon$.
Each test particle follows in
the region $U>0$ a trajectory with the inclination angle $\beta^+$ until it
reaches the spherical impulse at the point represented by $\alpha^+$.
The impulse influences the particle in such a way that it emerges
in the region $U<0$ at the point given by $\alpha^-$
and continues to move uniformly along the straight trajectory
with inclination $\beta^-$. Note that the lines in Fig.~2
represent just the inclination of the geodesic trajectories, not
the speed and orientation of the motion --- these will be
investgated later on.

From Fig.~2 it becomes obvious that the dependence of $\beta^-$
on the data $\alpha^+$, $\beta^+$, and the parameter $\epsilon$
is rather delicate. To shed some light on the details of
this dependence we introduce two special ``incoming'' inclination angles
denoted by $\beta^+_\parallel(\alpha^+)$ and $\beta^+_\perp(\alpha^+)$
defined by the property that the corresponding geodesic
{\it behind} the impulse are {\it parallel} ($\beta^-=0$) resp.\
{\it perpendicular} ($\beta^-=\pm\frac{\pi}{2}$)
to the strings localized along the $z$-axis.
It follows immediately from (\ref{beta}) that
\begin{eqnarray}
\cot\beta^+_\parallel&=&\frac{E-D}{2F}\ ,\label{parper}\\
\cot\beta^+_\perp&=& \frac{(1-\epsilon)\cot\alpha^+-(1-\delta)(E-D)\cot\alpha^-}
  {(1-\epsilon)-(1-\delta)\,2F\,\cot\alpha^-}\ ,\nonumber
\end{eqnarray}
where the functions $D,E,F$ are given by (\ref{EDF}), $Z_0$ by
(\ref{Z0}), and $\alpha^-$ by (\ref{alpha}). The functions (\ref{parper}) are
drawn in Fig.~3 for the three types of spacetimes given by $\epsilon=0, -1, +1$,
and for a ``typical'' value of the deficit-angle parameter $\delta=0.3\,$.
Both $\beta^+_\parallel$ and $\beta^+_\perp$ vanish at
$\alpha^+=0$. For $\epsilon=0$ the 
functions $\beta^+_\parallel(\alpha^+)<\alpha^+<\beta^+_\perp(\alpha^+)$
monotonically increase to the values
$\cot\beta^+_\parallel=(1-\delta)/[\delta(1-\frac{1}{2}\delta)]$,
$\beta^+_\perp=\frac{\pi}{2}$
at $\alpha^+=\frac{\pi}{2}$. For $\epsilon=-1$ the relation is
$\beta^+_\parallel>\beta^+_\perp>\alpha^+$, and the corresponding
values at $\alpha^+=\frac{\pi}{2}$ are $\beta^+_\parallel={\pi}$,
$\beta^+_\perp=\frac{\pi}{2}$. In the case $\epsilon=1$ these two
functions {\it coincide} for all values of $\alpha^+$, which directly
follows from (\ref{parper}). The functions grow to a maximum value
and then decrease to $\beta^+_\parallel=\beta^+_\perp=\frac{\pi}{2}$
at $\alpha^+=\frac{\pi}{2}$. The graphs presented in Fig.~3
provide a qualitative picture of the character of the
trajectories depending on the
choice of the initial angles $\alpha^+$, $\beta^+$.
Trajectories close to $\beta^+_\parallel(\alpha^+)$ become ``nearly
horizontal'' behind the impulse, whereas those  close to
$\beta^+_\perp(\alpha^+)$ become ``nearly vertical''. Since for
$\beta^+_\parallel$ we have $\dot x^-_0=0$ whereas
$\beta^+_\perp$ implies $\dot z^-_0=0$, it
follows that  the particles actually {\it stop}  behind
the impulse if we choose $\beta^+=\beta^+_\parallel(=\beta^+_\perp)$
for a given $\alpha^+$ in the impulsive spacetime with $\epsilon=+1$.

\begin{figure*}
\centering
\includegraphics[width=.75\textwidth]{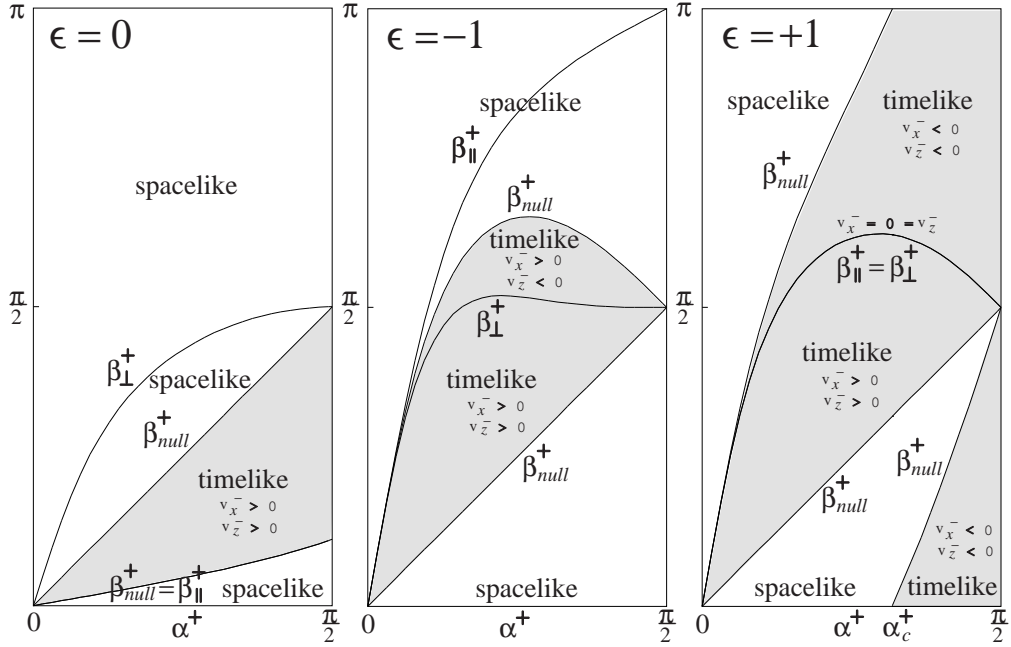}
\caption{\label{fig3}
Plots of $\beta^+_\parallel$, $\beta^+_\perp$ and
$\beta^+_{null}\,$, introduced in the text, as functions of the
angle $\alpha^+$ (again $\delta=0.3$).
The causal character of the geodesics with specific initial data
is indicated by the shading of the respective regions: grey resp.\
white corresponds to timelike resp.\ spacelike geodesics.
}%
\end{figure*}

Note, however, that the refraction formula (\ref{beta}) while
relating the {\it inclination} of the trajectories behind
the wave to its initial values does not provide any information on
the specific {\it speed} and {\it orientation} of the motion.
Of course, the velocity is
proportional to the parameters  $\dot x_0, \dot y_0, \dot z_0$,
which are the derivatives of space coordinates with respect to the
affine parameter $\tau$, see equation (\ref{geod+}). However, for
physical interpretation of the motion we need the {\it velocity with
respect to the Minkowski frame behind the impulse}, which is given
by
\begin{equation}
(v_x^-,v_y^-,v_z^-)=\left(\frac{\dot x_0^-}{\dot t_0^-},
\frac{\dot y_0^-}{\dot t_0^-},\frac{\dot z_0^-}{\dot t_0^-}\right)\ .
\end{equation}
From (\ref{explicit}) using (\ref{ab+}) we obtain
\begin{equation}\label{vxvz}
\begin{aligned}
v_x^- &= \frac{G}{{{\textstyle \frac{1}{2}}(Z_0+Z_0^{-1})\,G -(1+\epsilon)(\cot\alpha^+-\cot\beta^+)}} \ ,\\
v_z^- &= \frac{{{\textstyle \frac{1}{2}}(Z_0-Z_0^{-1})\,G-(1-\epsilon)(\cot\alpha^+-\cot\beta^+)}
}{ {{\textstyle \frac{1}{2}}(Z_0+Z_0^{-1})\,G -(1+\epsilon)(\cot\alpha^+-\cot\beta^+)}}\ ,
\end{aligned}
\end{equation}
and $v_y^-=0$, where $G=(1-\delta)[(E-D)- 2F\cot\beta^+]$. The above expressions
determine the velocity of the particle (including its orientation) in the region behind
the impulse as a function of the initial parameters $\alpha^+$, $\beta^+$:
The particle moves from the point
$\alpha^-(\alpha^+)$ in the refracted direction $\beta^-(\alpha^+,\beta^+)$
given by (\ref{alpha}) (\ref{beta}) in terms of the parameters
$\alpha^+$, $\beta^+$ and its velocity is given by (\ref{vxvz}).

Each geodesic belonging to the family (\ref{privgeod})
following the trajectory determined by $\alpha^+$, $\beta^+$ has a specific
causal character. If the magnitude of the velocity is such that
$v^-\equiv\sqrt{(v_x^-)^2+(v_z^-)^2}<1$, the geodesic is {\it timelike}
\hbox{($e=-1$)}. When $v^-=1$ it is {\it null} ($e=0$), and for $v^->1$ it is
{\it spacelike} ($e=+1$). Moreover, we can express the condition
for the null  geodesics explicitly. Substituting from
(\ref{vxvz}) we obtain a quadratic equation for $\cot\beta^+$
which can be solved. In the range $\alpha^+\in(0,\frac{\pi}{2}]$
there are always two real roots, namely
\begin{eqnarray}
1.&&\!\!\!\!  \beta_{null}^+=\alpha^+\ ,\label{betanull}\\
2.&&\!\!\!\!  \cot\beta_{null}^+=\frac{{\epsilon\cot\alpha^+
   -{\textstyle \frac{1}{2}}(Z_0^{-1}+\epsilon  Z_0)(1-\delta)(E-D)}}
 {{\epsilon -{\textstyle \frac{1}{2}}(Z_0^{-1}+\epsilon Z_0)(1-\delta)\,2F}} \ .\nonumber
\end{eqnarray}
The first equation in (\ref{betanull}) implies that {\it all geodesics of the
family} (\ref{privgeod}) {\it which move radially ``outside'' the impulse are
null}. In fact, it follows from (\ref{inver}) and (\ref{ab+}) that
$\dot U_0=0$, i.e., $U\equiv0$. These are exactly those null geodesics which {\it
generate the spherical impulse itself}.
Non-trivial null geodesics which cross the impulsive wave are
thus given by the second root in (\ref{betanull}). 
Fig.~3 shows the functions $\beta_{null}^+(\alpha^+)$ for the three
spacetimes characterized by $\epsilon=0$, $\epsilon=-1$, and
$\epsilon=+1$ respectively.

For $\epsilon=0$ it follows immediately from (\ref{betanull}) and
(\ref{parper}) that $\beta_{null}^+=\beta_\parallel^+$.
Therefore, these {\it null} geodesics are refracted to rays which
 are {\it parallel} to the strings behind the impulse ($v_x^-=0, v_z^-=1$).
All geodesics with trajectories given by $\alpha^+$, $\beta^+$ such
that $\beta^+_\parallel<\beta^+<\alpha^+$ are {\it timelike} with
$v_x^->0, v_z^->0$. All other geodesics are spacelike.

In the case $\epsilon=-1$ it can be shown that  the
non-trivial root $\beta^+_{null}$ in (\ref{betanull}) satisfies
the relation $\beta^+_\parallel>\beta^+_{null}>\beta^+_\perp>\alpha^+$,
as shown in Fig.~3. Again, in the region $\beta^+\in(\alpha^+,
\beta^+_{null}(\alpha^+))$ all the geodesics are timelike with $v_x^->0$.
At $\beta^+_\perp$ the velocity $v_z^-$ changes sign:
for $\beta^+<\beta^+_\perp$ we have $v_z^->0$, and for $\beta^+>\beta^+_\perp$
we obtain $v_z^-<0$.

\begin{figure*}
\centering
\includegraphics[width=.71\textwidth]{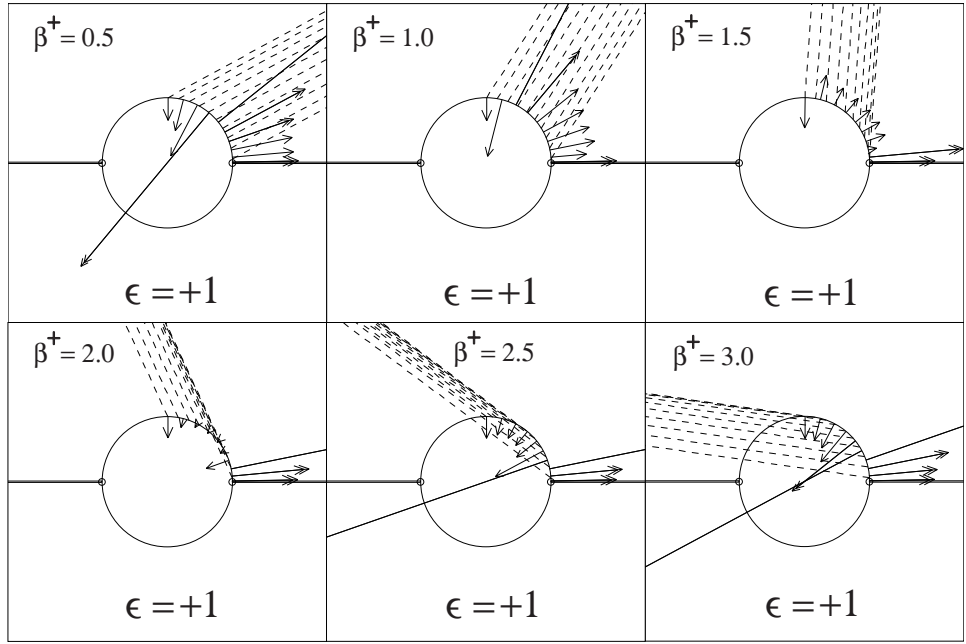}
\caption{\label{fig4}
In the region $U<0$ behind the expanding  spherical impulse
with $\epsilon=+1$, the motion of test particles is always
radial, i.e., exactly (de)focusing. Incoming trajectories for
various $\beta^+$ and $\alpha^+$ in the region $U>0$, with the
deficit angle parameter $\delta=0.3$, are indicated
by dashed lines. The velocity vectors behind the wave are
indicated by arrows of the corresponding length and orientation
(tachyons are denoted by double arrows).
}%
\end{figure*}

Finally, the most interesting case is $\epsilon=+1$. In this case the graph shown
in Fig.~3 is even more delicate. For a particular value $\alpha^+_c$ given by the
condition $\cot(\frac{1}{2}\alpha^+_c)=(\Delta+\sqrt{\Delta^2-1})^{1-\delta}$,
where $\Delta^{-2}=2\delta(1-\frac{1}{2}\delta)$, the denominator on the right
hand side of (\ref{betanull}) vanishes. Therefore, the value of
$\beta^+_{null}$ jumps at $\alpha^+_c$ from  $\beta^+_{null}(\alpha^+_{c-})=\pi$
to $\beta^+_{null}(\alpha^+_{c+})=0$. (Note that $\alpha^+_c$ monotonically increases
from $0$ to $\frac{\pi}{2}$ as the string parameter $\delta$ grows from the value
$0$ to $1$.) Consequently, there are {\it two disconnected regions
of timelike geodesics}. In the ``upper''
region ($\alpha^+<\beta^+<\beta^+_{null}$) lies the line
$\beta^+_\parallel=\beta^+_\perp$: particles moving along
timelike geodesics with the ``incoming position'' $\alpha^+$ and
a suitable ``inclination'' $\beta^+=\beta^+_\parallel(\alpha^+)=\beta^+_\perp(\alpha^+)$
will {\it exactly stop} at the point $x_0^-, z_0^-$ in the region behind the impulse
($v_x^-=0=v_z^-$), see (\ref{vxvz}). Particles below this boundary
($\beta^+<\beta^+_\parallel=\beta^+_\perp$) {\it move radially
outward} since $v_x^->0, v_z^->0$. On the other hand, timelike
particles with $\beta^+>\beta^+_\parallel=\beta^+_\perp$ for a
given $\alpha^+$ have $v_x^-<0, v_z^-<0$ so that these {\it
radially approach  the origin}. Thus there is an {\it exact focusing
effect of the impulse} on these timelike geodesics. The same is
true for all timelike geodesics in the ``lower region''
corresponding to initial values $\alpha^+$ close to $\frac{\pi}{2}$ and
small $\beta^+$ (see Fig.~3).  The time
$t^-_f=2x_0(\cot\alpha^+-\cot\beta^+)/(E-D-2F\cot\beta^+)$
when each individual particle in the spacetime
$\epsilon=+1$ reaches the origin, depends on the particular initial
data. These geodesics are explicitly
drawn in Fig.~4 with arrows indicating the precise value of the
particle velocity behind the impulsive wave. Double arrows correspond
to tachyons moving along spacelike geodesics.
Notice that for large values of $\beta^+$ and small  $\alpha^+$
some of the incoming trajectories (dashed lines) are drawn inside
the circle which indicates the impulse. However, this is not a
contradiction as the figure represents just a snapshot at a given
time. In fact, the corresponding timelike particles move in the
outer region $U>0$ until they are hit by the expanding impulse 
(which at previous instants of time is a smaller circle). The 
tachyons move  "acausally" and thus their motion is neither intuitive nor
represents the motion of a test particle; this case is included for the 
sake of completeness.

All the above results can equivalently be obtained also by the
``inverse approach'', i.e., starting with the initial data (\ref{geod+b}) behind
the impulse (in the region $U<0$) and evolving these ``backward''
in time into $U>0$. The solution is given by
(\ref{cone}) which has to be substituted into (\ref{coef++}). Let
us  demonstrate this method by considering a simple yet important
particular example. Consider a geodesic motion of {\it timelike}
particles which are {\it at rest behind} the impulse generated by
a snapping cosmic string, $\dot x_0=\dot y_0=\dot z_0=0$ (so that
$\gamma=1$). In other words, we investigate motion of those particles
which are exactly stopped by the impulse. Again, we can
without loss of generality assume  that $y_0=0$. Then the
constraint (\ref{constraint}) implies $\epsilon=+1$ so that such
a situation  may occur only in the spacetime with this value
of the parameter $\epsilon$. Relations (\ref{cone}) then immediately
yield
\begin{equation}
Z_0=\frac{x_0}{\tau_i-z_0},\
V_0=\sqrt2 \tau_i,\
\dot V_0=0,\
\dot U_0=-\frac{1}{\sqrt2}\ .\label{spec}
\end{equation}
The motion of the particles in front of the wave is thus given by
(\ref{coef++}) with the parameters substituted from  (\ref{spec})
and (\ref{ABCDEF}). We obtain
\begin{eqnarray}
&&x^+_0 +\, \hbox{i}\, y^+_0=2\tau_i C\ ,\quad
 \dot x^+_0 +\, \hbox{i}\, \dot y^+_0=F\ , \nonumber\\
&& z^+_0=\tau_i(B-A)\ , \quad\
 \dot z^+_0={\textstyle\frac{1}{2}}(E-D)\ , \label{speca}\\
&& t^+_0=\tau_i(B+A)\ ,\quad\ \
 \dot t^+_0={\textstyle\frac{1}{2}}(E+D)\ . \nonumber
\end{eqnarray}
As expected, $y^+_0=0=\dot y^+_0$ since the coefficients $C$ and
$F$ are real. From the remaining relations we easily derive
(using the definitions (\ref{parper}) and the fact that $\epsilon=+1$)
\begin{eqnarray}
&&\cot\alpha^+ \equiv \frac{z^+_0}{x^+_0}=\frac{B-A}{ 2C}
  ={\textstyle\frac{1}{2}}(Z_0^{1-\delta}-Z_0^{\delta-1})\ ,\nonumber\\
&&\cot\beta^+  \equiv \frac{\dot z^+_0}{\dot x^+_0}=\frac{E-D}{2F}
  \equiv\cot\beta_\parallel\equiv\cot\beta_\perp\ . \label{specb}
\end{eqnarray}
Of course, these results are identical to those obtained previously
using the ``direct'' approach. However, now we know {\it
explicitly} how to choose the initial data $\alpha^+$, $\beta^+$
to put the particle at rest behind the impulse at time $\tau_i$ in
the specific point $x_0$, $z_0$. For this, one simply substitutes
$Z_0=x_0/(\tau_i-z_0)$ into (\ref{specb}).

\subsection{General $C^1$-geodesics}

Let us recall again that all the  geodesics in the
spacetime (\ref{en0}) with the impulsive gravitational wave generated
by a snapping cosmic string (\ref{string}) which we have investigated so
far, are very special, i.e., $Z=Z_0=$const.\ (cf.\ (\ref{privgeod})).
They are geometrically preferred since they
are {\it restricted to a single plane} (taken above as $y=0$)
which also contains the snapping string localized along the $z$-axis. This fact
immediately follows from the constraint (\ref{constraint}).
Therefore, the corresponding particles move --- although not necessarily
parallel --- ``along'' the strings. To investigate  more
general geodesics which ``bypass'' the strings, we have to relax the
condition $Z=Z_0$. However, these general geodesics
with $\dot Z_0\not=0$  cannot be found easily in the continuous
form of the metric (\ref{en0}). Nevertheless, in (\ref{const})-(\ref{complcoef++})
we presented an {\it explicit form of general geodesics} which was derived under the
assumption that these are $C^1$ in the continuous coordinate system (\ref{en0}).

As an interesting particular example, which can be investigated using these
expressions, let us now consider geodesics in the $z=0$ plane only. This
is the plane of symmetry {\it  perpendicular to the strings}. We assume that
$z_0=0=\dot z_0$ in the region $U<0$
behind the wave. With this, the relations (\ref{genercont}) simplify to
\begin{eqnarray}
&&Z_i=\frac{x_0 +\, \hbox{i}\, y_0}{ \gamma\tau_i}\ ,\qquad\quad
  V_i={\textstyle\frac{1}{\sqrt2}}\,(1+\epsilon)\gamma\tau_i\ ,\nonumber\\
&&\dot Z_i=\frac{\dot x_0 +\, \hbox{i}\, \dot y_0}{ \gamma\tau_i} \nonumber\\
&&\hskip9mm-(x_0 +\, \hbox{i}\, y_0)\frac{(1-\epsilon)\gamma^2\tau_i +2\epsilon(x_0\dot x_0+y_0\dot y_0)
}{(1+\epsilon) (\gamma\tau_i)^3} \ ,\nonumber\\
&& \dot V_i=\frac{(1-\epsilon)^2\gamma^2\tau_i+4\epsilon(x_0\dot x_0+y_0\dot y_0)
  }{ \sqrt2(1+\epsilon) \gamma\tau_i}\ ,\nonumber\\
&&\dot U_i={\textstyle{\sqrt2}}\,\,
  \frac{x_0\dot x_0+y_0\dot y_0-\gamma^2\tau_i}
  {(1+\epsilon)\gamma\tau_i}\ ,\label{genercontsp}
\end{eqnarray}
from which follows that $|Z_i|=1$. Therefore 
the coefficients (\ref{ABCDEFgener}) entering (\ref{complcoef++}) 
take the following form
 \begin{eqnarray}
&&A=B= \frac{1}{(1-\delta)(1+\epsilon)}\ ,\
C= \frac{Z_i^{1-\delta}}{(1-\delta)(1+\epsilon)}\ ,   \nonumber\\
&&D= \frac{(\textstyle{\frac{1}{2}\delta})^2
    +\epsilon(\textstyle{1-\frac{1}{2}\delta})^2}{1-\delta}\ ,\
  E= \frac{(\textstyle{1-\frac{1}{2}\delta})^2
    +\epsilon(\textstyle{\frac{1}{2}\delta})^2}{1-\delta}\ ,\nonumber\\
&&F= \frac{ Z_i^{1-\delta}\textstyle{ \frac{1}{2}\delta(1-\frac{1}{2}\delta)(1+\epsilon)}}
   {1-\delta}\ ,\label{ABCDEFspec}\\
&&A_{,Z}= \frac{\textstyle{\frac{1}{2}\delta}
    -\epsilon(\textstyle{1-\frac{1}{2}\delta})}{(1-\delta)(1+\epsilon)^2Z_i}\ ,\
  B_{,Z}= \frac{(\textstyle{1-\frac{1}{2}\delta})
    -\epsilon\textstyle{\frac{1}{2}\delta}}{(1-\delta)(1+\epsilon)^2Z_i}\ ,\nonumber\\
&&C_{,Z}=  \left(\frac{\bar Z_i}{Z_i}\right)^{\delta/2}\frac{(\textstyle{1-\frac{1}{2}\delta})
    -\epsilon\textstyle{\frac{1}{2}\delta}}{(1-\delta)(1+\epsilon)^2}\ ,\nonumber\\
&&C_{,\bar Z}=  \left(\frac{\bar Z_i}{Z_i}\right)^{\delta/2-1}
  \frac{\textstyle{\frac{1}{2}\delta}-\epsilon(\textstyle{1-\frac{1}{2}\delta})}
    {(1-\delta)(1+\epsilon)^2}    \ .\nonumber
 \end{eqnarray}
Substituting (\ref{genercontsp}), (\ref{ABCDEFspec}) into
(\ref{complcoef++}) we obtain an explicit solution which describes the
behavior in the region $U>0$ outside the impulse. In particular, we  easily derive that
\begin{eqnarray}
&&z^+_0\equiv{\textstyle\frac{1}{\sqrt2}}(\CU_0^+-\CV_0^+)=0\ ,\nonumber\\
&&\dot z^+_0\equiv{\textstyle\frac{1}{\sqrt2}}(\dot\CU_0^+-\dot\CV_0^+)=0\ .\label{e1}
\end{eqnarray}
Therefore, the geodesics {\it remain in the plane} $z=0$ also
in the outside region, as is expected from the symmetry of the
spacetime. Straightforward but somewhat lengthy calculations 
for
$\eta^+_0\equiv\frac{1}{\sqrt2}(x^+_0+\,\hbox{i}\,y^+_0)$,
$\dot\eta^+_0\equiv\frac{1}{\sqrt2}(\dot x^+_0+\,\hbox{i}\,\dot  y^+_0)$
yield
\begin{eqnarray}
&&x^+_0+\,\hbox{i}\,y^+_0=\frac{(\gamma\tau_i)^\delta}{1-\delta}\,(x_0+\,\hbox{i}\,y_0)^{1-\delta}\ ,\label{e2}\\
&&\dot x^+_0+\,\hbox{i}\,\dot  y^+_0=\left(\frac{x_0-\,\hbox{i}\,y_0}{ x_0+\,\hbox{i}\,y_0 }\right)^{\delta/2}
   \nonumber\\
&&\hskip14mm\times \,\Big[(\dot x_0 +\, \hbox{i}\, \dot y_0)-{\cal P}(x_0 +\, \hbox{i}\, y_0)\Big]\ ,\label{e3}
\end{eqnarray}
where
\begin{equation}
{\cal P}= -\frac{\delta}{ 1-\delta}
\left(\frac {{\textstyle\frac{1}{2}}\delta(x_0\dot x_0+y_0\dot y_0)}{ x_0^2+y_0^2}
  + \frac{1-{\textstyle\frac{1}{2}\delta}}{\tau_i}\right)\ ,\label{P}
\end{equation}
and $\tau_i=\sqrt{(x_0^2+y_0^2)/(\dot x_0^2+\dot y_0^2-e)}$.
The equations  (\ref{e2}) and (\ref{e3}) 
describe the identification of points on the
impulse, and the refraction formula in the transverse plane $z=0$,
respectively. These admit a natural geometrical interpretation. If
we introduce a ``polar'' representation of positions and
velocities by $x_0 +\, \hbox{i}\, y_0\equiv\rho_0\exp(\hbox{i}\phi_0)$,
$\dot x_0 +\, \hbox{i}\, \dot y_0\equiv\dot
\rho_0\exp(\hbox{i}\dot\phi_0)$, we can conclude from (\ref{e2}) that
$\phi_0^+=(1-\delta)\phi_0$. As the range of $\phi_0$ inside (behind) the
spherical impulse spans the whole Minkowski space,
$\phi_0\in[0,2\pi)$, the range of the angular parameter $\phi^+_0$
outside is $[0,2\pi(1-\delta))$. Therefore, there is a {\it
deficit angle } $2\pi\delta$ in front the impulse corresponding to
the presence of the (snapped) cosmic string. This is in full
agreement  with the geometrical construction of the spacetime
presented e.g., in \cite{PodGri00}. The relation (\ref{e3}) is the
{\it refraction formula} for geodesics in the symmetry plane  $z=0$
perpendicular to the strings. Interestingly, here the effect is
totally {\it independent} of the parameter $\epsilon$, i.e., the
differences between the spacetimes characterized by
$\epsilon=0,-1,+1$ disappear in this plane of symmetry. Of course, for
$\delta=0$  we obtain a trivial solution $\dot y^+_0=\dot y_0$, $\dot x^+_0=\dot
x_0$ in the complete Minkowski space without 
string and impulse. Note also that the factor
\begin{equation}
\left(\frac{x_0-\,\hbox{i}\,y_0}{ x_0+\,\hbox{i}\,y_0 }\right)^{\delta/2}
\equiv\>\exp(-\hbox{i}\delta\phi_0)
\end{equation}
in (\ref{e3}) is just an appropriate ``rectifying''complex unit factor
which ensures the one-to-one correspondence between the identified
points on both sides of the impulse (analogously to the function
$\alpha^-(\alpha^+)$ given by (\ref{alpha}) for longitudinal motion).
This can be seen easily if we consider two infinitesimally close
parallel null geodesics $y_0=0=\dot y_0$ in the Minkowski region
$U<0$ without topological defects. The first geodesic is given by
$\phi_0=0$, the second one by an angle $\phi_0$ near $2\pi$. However, from the formula
(\ref{e3}), which reads $\dot\rho^+_0\exp(\hbox{i}\dot\phi^+_0)={\cal
F}\exp(-\hbox{i}\delta\phi_0)$, where ${\cal F}$ is a real factor,
it follows that $\dot\phi^+_0=-\delta\phi_0$. Therefore, outside
the impulse the two geodesics which {\it remain parallel} are
described by $\dot\phi^+_0=0$ and $\dot\phi^+_0$ near to $-2\pi\delta$,
respectively. The difference $2\pi\delta$ exactly corresponds to the
deficit angle in the (locally) flat space with the string outside the
spherical impulse. Therefore, the ``pure'' physical refraction effect
of the impulse on geodesics is described just by the expression in
the square bracket on the right hand side of the equation (\ref{e3}).

The above relations can easily be applied to investigate the effect
of the impulsive  wave on a {\it ring of free test particles}. Let us
consider a ring in the $z=0$ plane, centered around $x=0=y$,
consisting of particles  which are {\it at rest in front} of the wave,
$\dot x^+_0=0=\dot y^+_0$, in the (locally) flat Minkowski region $U>0$.
All the particles are simultaneously hit by the impulse at the
instant $\tau_i$ and the ring starts to deform according to (\ref{e2}),
(\ref{e3}). Obviously, it follows from (\ref{e3}) that the velocities of the
particles $\dot x_0$, $\dot y_0$ behind the impulse ($U<0$) are
given by
\begin{equation}
\dot x_0={\cal P}\,x_0\ ,\quad
\dot y_0={\cal P}\,y_0\ ,\label{prop}
\end{equation}
with ${\cal P}$
given by (\ref{P}) and $\tau_i^{-1}=\sqrt{{\cal P}^2+(x_0^2+y_0^2)^{-1}}$.
This yields a self-consistent solution only if
\begin{equation}
{\cal P}= -\frac{\delta(1-{\textstyle\frac{1}{2}\delta}) }{(1-\delta)
\sqrt{x_0^2+y_0^2}} \ .\label{ring}
\end{equation}
Thus, all the particles of the ring move {\it radially}
towards the origin in the $z=0$ plane, with the {\it same velocity}
$v\equiv\sqrt{\dot x_0^2+\dot y_0^2}
=\delta(1-{\textstyle\frac{1}{2}\delta})/(1-\delta)$. The ring is
deformed by the impulse into {\it contracting and concentric circles}. Of course,
this is in accordance with the axial symmetry of the spacetime.

A more general situation in which the impulse deforms a {\it sphere}
of test particles (around the origin) initially at rest is, however,  more
difficult to investigate explicitly. We can again employ the coordinate freedom
$Z\to Z\,\exp(\hbox{i}\phi)$ related to the axial symmetry
of the spacetime which corresponds to a simple rotation of the
$(x,y)$-planes around the $z$-axis. Using (\ref{ABCDEFgener})  and
(\ref{complcoef++}) we conclude
$\eta_0^+\to\eta_0^+\,\exp[\hbox{i}(1-\delta)\phi]$,
$\dot\eta_0^+\to\dot\eta_0^+\,\exp[\hbox{i}(1-\delta)\phi]$.
Therefore, without loss of generality we can always set for each
{\it individual} test particle $\eta_0^+$ to be real, i.e., $y_0^+=0$.
Moreover, we are considering the motion of
test particles which are {\it at rest outside} the expanding impulsive
wave, $\dot\eta_0^+=0$, $\dot z_0^+=0$. From (\ref{complcoef++}), (\ref{ABCDEFgener})
and (\ref{genercont}) it then follows that $Z_i$ and $\dot Z_i$ are
real so that $y_0=0=\dot y_0$. The sphere of test particles is thus
deformed into an {\it axially symmetric} surface 
which is fully described by the section $y=0$.

Setting $y_0=0=\dot y_0$ in (\ref{genercont}) we can now simplify
$\dot Z_i$ to
\begin{eqnarray}
&&\dot Z_i= \label{dotZ}\\
&&\   \frac{[(1-\epsilon) \gamma\tau_i-(1+\epsilon)z_0]\dot x_0 -
 [(1-\epsilon)\gamma-(1+\epsilon)\dot z_0]x_0 }
 {(\gamma \tau_i-z_0)[(1+\epsilon) \gamma\tau_i-(1-\epsilon)z_0]} \ .
\nonumber
\end{eqnarray}
Consequently, for these geodesics $\dot Z_i=0$ if and only if the
constraint (\ref{constraint}) is satisfied. In such a case, the
geodesics reduce to the privileged family (\ref{privgeod}) for which
$Z_i=Z_0=\,$const.\ which we investigated in detail above.
However, these special geodesics {\it exclude} observers which are
static in the Minkowski region outside the impulse. Indeed, from the conditions $\dot x_0^+=0=\dot
z_0^+$ we obtain using (\ref{complcoef++}) the relation
$(A-B)F=(D-E)C$. Substituting from (\ref{ABCDEF}) this reduces to
 $Z_0^{\delta-1}(\frac{1}{2}\delta p-\epsilon Z_0^2)=
Z_0^{1-\delta}(\frac{1}{2}\delta p-1)$,
which has no solution except for observers in the plane $z=0$ in
spacetime with $\epsilon=+1$, which we investigated in
(\ref{prop}), (\ref{ring}).

Therefore, to obtain a nontrivial family of geodesics
corresponding to initially static test particles, one has to
consider the more complicated situation in which $\dot Z_i\not=0$. It is
difficult to obtain the description of these geodesics in an explicit
form. Nevertheless one can immediately argue that the motion can
not be spherically symmetric. For example, for the case
$\epsilon=+1$ we observe from (\ref{dotZ}) that $z_0\dot x_0\not =x_0\dot z_0$
which, in terms of (\ref{ab-}), can be expressed as
$\alpha^-\not=\beta^-$. Obviously, the trajectories of such
geodesics behind the impulse {\it are not radial}, i.e., these do not
``point'' towards the origin. A sphere of free test particles
which are at rest in the Minkowski region outside the expanding impulsive
wave is thus not deformed into spherical shapes, but to a more
complicated (axially symmetric) surface.

\section{Concluding remarks}

We  presented a complete solution of geodesic motion --- although
not always in closed explit form ---
which describes the effect on free particles of expanding spherical impulsive
gravitational waves propagating in a flat background.
In particular, we discussed in detail the geodesics
in the axially symmetric spacetimes with the impulse
generated by a snapping cosmic string. The above results can be
used not only for physical interpretation of the behavior of free test particles
but also as a starting point for a mathematically
rigorous distributional treatment of impulsive Robinson--Trautman
spacetimes. To be more specific, the geodesics of the special family (\ref{privgeod})
provide the key to understand the discontinuous transformation relating
the distributional and the continuous form or the metric (analogous
to the case of impulsive pp-waves; cf.\ \cite{KunSt99}).
These interesting questions will, however, be investigated elsewhere.

\acknowledgments

The present work was supported,
in part, by the grant GACR-202/02/0735 of the Czech Republic
and grant P-12023MAT of the Austrian Science Found.

\end{document}